# AI-Generated Letters from the Future:
# A Randomized Test of Personalized Climate Communication

Nattavudh Powdthavee[1*], Pat Pataranutaporn[2], Sandra J. Geiger[3], Louisa Richter[4], Mathew P. White[4]

[1] *Nanyang Technological University, Singapore*
[2] *Massachusetts Institute of Technology*
[3] *Princeton University*
[4] *University of Vienna*

**Abstract**

We examined whether personalized, AI-generated letters from the future can increase public engagement with climate action. In a preregistered online experiment with 1,654 U.S. parents, participants were randomly assigned to receive either a fact-based climate report, an AI-generated letter from a generic future person, or an AI-generated letter framed as written by their future child. Although both narrative conditions increased empathic concern for future generations, neither had a detectable effect on stated climate policy support or donations to an environmental charity. Personalizing the message as coming from one's future child did not enhance its impact. Exploratory analyses suggest that both narratives led to more emotionally differentiated appraisals of future scenarios, yet also made desirable climate outcomes seem less likely. These findings highlight key constraints on the effectiveness of AI-generated narrative interventions and underscore the importance of balancing emotional resonance with perceived credibility in climate communication.

[*] Corresponding author: Nanyang Technological University, 48 Nanyang Avenue, Singapore, 639818. Email: nick.powdthavee@ntu.edu.sg. We have secured ethical approval from the Institutional Review Board at the Massachusetts Institute of Technology (MIT IRB Exempt ID: E-6515). NP gratefully acknowledges financial support from NTU's Start-up Grant (022674-00001). Competing interests: The author(s) declare none.

# 1. Introduction

Despite scientific consensus and rising public concern, climate change still demands coordinated responses at both systemic and individual levels (Gifford, 2011; Norgaard, 2011). Yet current responses remain inadequate in scale and effectiveness (OECD, 2024), and behavioural change continues to play a crucial role in shaping public support for policy and complementary individual action. As a result, communication-focused interventions such as social norm nudges, carbon labelling, and default green energy plans have been deployed to encourage climate-friendly choices (Allcott & Rogers, 2014; Carlsson et al., 2021). However, these approaches often produce modest, context-dependent effects that are sensitive to implementation and audience characteristics. Given the centrality of public engagement to successful climate mitigation (Loewenstein & Chater, 2017; Zhao & Chen, 2023) and evidence that many existing interventions achieve limited impact (Vlasceanu et al., 2024), identifying more effective strategies for motivating climate action remains a pressing challenge.

A substantial body of research suggests that climate change is often perceived as a distant threat, which can reduce feelings of urgency and engagement (Spence et al., 2012; Brügger et al., 2016). Psychological distance—the subjective perception of how removed something is from one's direct experience across temporal, spatial, social, and hypothetical dimensions (Trope & Liberman, 2010)—has been linked to lower concern and a lower willingness to act (Milfont, 2010; Gifford, 2011; Spence et al., 2012; Vlasceanu et al., 2024), and numerous interventions have sought to reduce this distance by localizing impacts or emphasizing immediacy. However, as recent reviews show, findings are highly inconsistent. Keller et al. (2022) conclude that psychological distance is conceptualized and measured inconsistently across studies, complicating the development of generalizable strategies. Maiella et al. (2020) report that lower levels of psychological distance are generally associated with greater pro-environmental engagement, yet some studies find reverse or null effects. Van Valkengoed et al. (2023), for instance, argue that most people perceive climate change as real and immediate, and that perceiving it as distant does not necessarily reduce pro-environmental action. These findings challenge the widely held assumption that interventions aimed to reduce psychological distance—particularly the temporal and spatial dimensions—can reliably promote climate action. Overall, psychological distance is a multifaceted and context-dependent construct and reducing temporal or spatial distance alone may be insufficient to motivate climate action.

This has prompted growing interest in more emotionally engaging strategies that have shifted from abstract or statistical framings toward more emotionally vivid and immersive formats, which are considered more likely to evoke empathy and moral engagement (McDonald et al., 2015; Slovic, 2007). For instance, virtual reality simulations that allow participants to experience climate impacts from a first-person perspective have been shown to heighten concern and reduce psychological distance (Thoma et al., 2023; Chatterjee et al., 2024; Pi et al., 2025). Other intergenerational strategies, such as letter-writing to future generations or climate conversations with children, have increased feelings of legacy, moral continuity, and concern for future others (Lawson et al., 2019; Syropoulos et al., 2023; Vlasceanu et al., 2024). These approaches underscore the value of interventions that help individuals vividly imagine the lived experiences of others, especially future generations.

Within intergenerational approaches, kinship-based personalization—emphasizing the consequences for one's own children rather than for distant others—appears particularly effective at motivating climate action. Fornwagner and Hauser (2022) found that parents invested substantially more in voluntary climate action when their child, rather than a stranger's child, was the observer. Syropoulos et al. (2023) similarly showed that writing a letter to one's future child increased legacy motivation and pro-environmental donations compared with general messaging. This kinship advantage extends beyond climate contexts: Children can meaningfully influence their parents' environmental attitudes, and family climate discussions are more effective than generic appeals (Lawson et al., 2019). A recent 63-country megastudy further identified future-self-continuity interventions and letters to future generations as among the most successful strategies for increasing climate-policy support (Vlasceanu et al., 2024). However, despite their promise, these approaches face scalability constraints—requiring substantial resources (e.g., VR), interpersonal dynamics (e.g., conversations with children), or facilitation (e.g., letter-writing)—which limit their feasibility for widespread deployment.

Recent advances in generative artificial intelligence (AI) offer a scalable and cost-effective way to produce emotionally resonant, personalized narratives. Beyond personalization, these tools can construct two coherent alternative futures—one positive and one negative—allowing individuals to imagine contrasting climate trajectories rather than a single deterministic outcome. This dual-future structure enables more vivid contrastive thinking, which may heighten emotional engagement while preserving narrative coherence. Early applications have shown promise: For example, conversational AI systems have increased environmental behavioral intentions relative to static educational formats (Pataranutaporn et al., 2025), and AI-generated future-self avatars have strengthened emotional connection to one's future identity (Hershfield et al., 2012; Pataranutaporn et al., 2024). We extend this work here by shifting from self-focused communication to an intergenerational narrative, framing climate futures as written by a participant's future child. This approach is expected to particularly reduce the social distance to climate change—a key dimension of future-oriented moral engagement (Liberman & Trope, 2008; Pahl & Bauer, 2013). Yet despite the intuitive appeal of personalized intergenerational AI narratives, their behavioral effects have not been rigorously tested at scale.

To address this gap, we conducted a preregistered online experiment designed to test the effects of kinship-based personalization in climate communication using AI. Our final sample consisted of 1,654 U.S. parents whose youngest child was aged 15 or below. Participants were randomly assigned to receive one of three messages: a fact-based climate report, an AI-generated letter from a future stranger, or an AI-generated letter from their future child. Each narrative letter included two alternative climate trajectories—one positive and one negative—leveraging AI's capacity to present vivid, parallel futures with narrative coherence. We then assessed support for a climate policy package, including the individual policies, and parents' incentive-compatible donation decisions, alongside hypothesized mechanisms intended to capture emotional, cognitive, and motivational responses to the letters.

We preregistered six hypotheses predicting that AI-generated narrative letters would increase climate policy support (Hypothesis 1) and donation behavior (Hypothesis 2) relative to a fact-based report, and that personalizing these letters as coming from participants' future children would strengthen these effects (Hypotheses 3 and 4). We also expected that presenting both a positive and a negative climate future would increase the emotional vividness and perceived immediacy of future outcomes. We further test the potential mechanisms underlying both interventions, including reducing psychological distance and increasing emotional responses (e.g., fear, hope, anticipated guilt) as well as motivation for climate action (Research Questions 1-4).

**Methods**

**2.1. Participants and sampling plan**

Following our pre-registered plans (https://osf.io/uq2w9/?view_only=c9fff96f26544c5d8f1c874a9b9b41ee), we recruited 3,000 U.S.-based participants via Prolific, an online research platform known for high-quality behavioral data (Peer et al., 2017). Eligibility criteria required participants to reside in the United States, be fluent in English, and have at least one child aged 0–15 years. After applying preregistered exclusion criteria—including two attention checks and data quality filters (i.e., likely bots, duplicates, fraudulent responses, as per *Qualtrics* recommendations)—we retained 2,422 participants.

As preregistered, participants were required to click through and read both future climate scenarios for the intervention to be delivered as intended. A total of 1,654 participants (68.3%) met this exposure criterion, while the remaining 31.7% either read only one of the two scenarios or did not read any of the letters. The final sample consisted of 659 participants in the control condition (39.8%), 515 in the generic AI condition (31.1%), and 480 in the personalized AI condition (29.0%). Restricting the sample to participants exposed to both future scenarios allows us to evaluate the intervention's effects among those who received the intended treatment.

Our sampling plan was based on an a priori power analysis using G*Power (version 3.1). For the main regression model—including two dummy-coded treatment indicators and no additional covariates—we estimated that a sample of 3,000 participants would provide 95% power to detect a small effect size of $f^2 = 0.00515$ (equivalent to Cohen's $d \approx 0.14$) at a 5% significance level. A secondary analysis comparing the two treatment groups (personalized vs. generic AI) would also be powered at 95% to detect an effect size of $f^2 = 0.00433$ (Cohen's $d \approx 0.13$). Additionally, with 1,000 participants per group, the study was powered at 84.8% to detect a five percentage-point difference in binary outcomes such as donation behavior, assuming a base rate of 20% and a one-sided logistic regression.

Although the final sample was smaller than anticipated, a post-hoc sensitivity analysis showed that with $n$ = 1,654, the study retained 73% power to detect very small effects ($f^2$ = 0.005), 96% power to detect small effects ($f^2$ = 0.01), and >99.9% power for small-to-medium effects ($f^2$ = 0.02). The design, therefore, remained robust to detecting practically meaningful differences, although caution is advised when interpreting null effects at very small magnitudes.

## 2.2. Experimental design and procedure

At the beginning of the experiment, all participants provided informed consent. During this process, the study was presented as "Perspectives on Future Scenarios" to minimize self-selection bias and avoid priming participants toward climate-related responses. After providing informed consent, participants completed an attention check to ensure that responses were not generated by automated bots. Following this, participants provided demographic information, including their child's age, their own age, gender, marital status, annual household income before tax, education level, ethnicity, and political identity.

Participants were randomly assigned to one of three experimental conditions, each presenting two contrasting climate scenarios for the year 2050. These were based on projections from the En-ROADS Climate Solutions Simulator and contrasted a future shaped by limited climate action (SSP2: Middle of the Road) with one shaped by proactive sustainability policies (SSP1: Sustainability). Both scenarios are based on data from the En-ROADS Climate Solutions Simulator (https://www.climateinteractive.org/en-roads/). The mean completion time was 14.20 minutes (*SD=7.67 minutes)*.

In the control condition, participants read a neutrally framed policy document summarizing each scenario in formal, evidence-based language. The "Future Person" condition presented the same content in the form of a narrative letter generated by an AI, written in a conversational tone and emphasizing lived experiences. In the "Future Child" condition, participants provided their child's name, age, gender, and location. They then selected an avatar that best represented their imagined child. The AI used this information to generate a tailored letter from the perspective of their future child in 2050, reflecting on life under each scenario. This design sought to explore whether personalization would reduce psychological distance and increase engagement.

All materials were delivered through an online Qualtrics survey, and the exact wording of each condition is provided in the additional details of the experimental procedures in Appendix A.

## 2.3. Primary outcomes

We preregistered two primary outcomes and one exploratory outcome. The first primary outcome was support for the entire climate policy package, as shown in Figure 1:

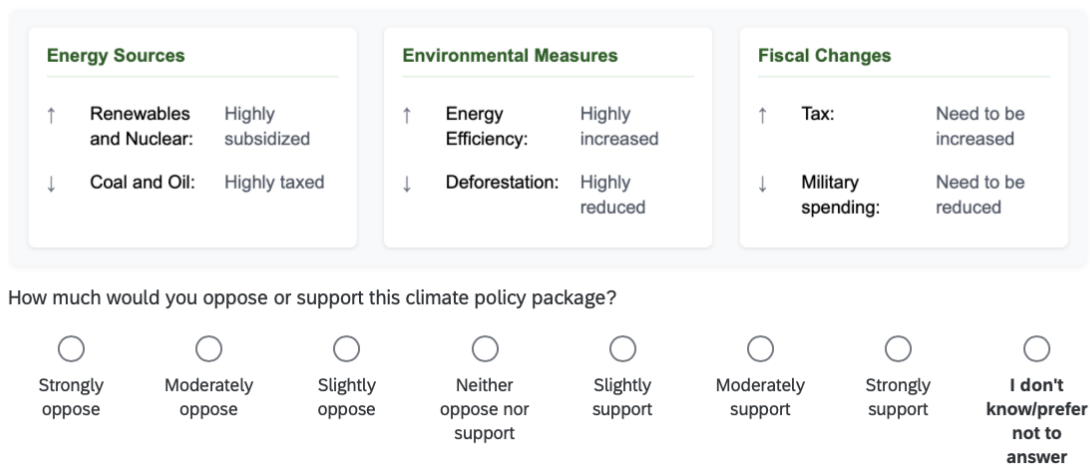

**Figure 1**: Primary outcome measure of support for the policy package

For exploratory purposes, participants also rated their support for the eight policy components, including four energy-related measures (highly taxed coal, highly taxed oil, highly subsidized renewables, and highly subsidized nuclear), two environmental measures (highly increased energy efficiency and highly reduced deforestation), and two fiscal changes (increased income tax and decreased military spending). The energy-related and environmental measures were assessed with the following item: "How much would you oppose or support each element of the climate policy package?", on 7-point Likert scales (from "Strongly oppose" to "Strongly support", with an additional "I don't know/prefer not to answer" option). Subsequently, participants were informed that the U.S. government would need to raise current income taxes and reduce military spending to adopt this climate policy package, and subsequently indicated their preferred income tax rate and preferred allocation to the military budget using slider scales from 0 to 100 (with the starting value set to the current average tax rate of 15% and the current U.S. military budget to 16%, respectively): "To adopt this climate policy, the US government will need to raise current income taxes, which will impact your financial situation. With this in mind, what income tax rate would you support? The current average income tax rate (2021) is around 15%. To put it simply, people in the US pay, on average, $15 in income tax for every $100 they earn. Please use the slider below to indicate a value. Values above 15% mean that you support a rise in income tax; values below 15% mean that you support a reduction in income tax" and "To adopt this climate policy, the US government will also need to reduce military spending. With this in mind, what military budget would you support? The current US military budget (2025) is 16% of the total federal budget. To put it simply, around $16 per $100 is spent on the military. Please use the slider below to indicate a value. Values above 16% mean that you support an increase in military spending; values below 16% mean that you support a reduction in military spending.

The second primary outcome was an incentive-compatible decision to donate. At the end of the study, participants were informed that they would receive a $0.50 bonus and could choose to keep it, donate it to a reforestation charity (*One Tree Planted*) or do neither. For analysis, we coded donation choices as a binary variable: 1 if the participant donated and 0 otherwise.

**2.4. Psychological mediators**

In addition to the outcomes, we collected a set of exploratory psychological-process measures to assess how participants engaged with the narratives. These included the following constructs in fixed order: perceived likelihood and valence of each of the scenarios, perceived psychological distance (adapted from Jones et al., 2017), perceived realism/vividness of the future scenario (Pataranutaporn et al., 2024), empathic engagement with future others (Lee & Li, 2023), and self-reported emotional responses to climate change. Perceived likelihood and valence of both scenarios described as "the future where the public majority rejects/supports the climate policy" were measured with self-constructed items: "How likely or unlikely you think it is that each of the two futures described will happen?" and "How good or bad do you think the two reported futures are?", using 7-point scales from *Extremely unlikely/bad* to *Extremely likely/good* (and an "I don't know/prefer not to answer" option).

Psychological distance was measured with two items, capturing the social and hypothetical dimensions: "Climate change will mostly affect people I do not know in 2050" and "I am uncertain what the effects of climate change will be in 2050" (1 *Strongly disagree* to 7 *Strongly agree*, with an additional "I don't know/prefer not to answer" option). Perceived realism/vividness of the future scenario was measured using the following question: "Does the child's letter seem realistic? (1 *Very unrealistic* to 7 *Very realistic*). Lastly, we also captured participants' emotional responses to climate change: "When you think about climate change and everything that you associate with it, how strongly, if at all, do you feel each of the following emotions?" (1 *Does not describe my feelings* to 5 *Clearly describes my feelings*; "I don't know/prefer not to answer"), including relaxed, hopeful, guilty, fearful, and motivated to take action in a randomized order.

Full descriptive statistics are reported in Appendix B.

## 2.5. Statistical analysis plan

All analyses followed our preregistered plan. Consistent with that plan, the primary analyses focus on participants who viewed both future scenarios, reflecting the intervention as delivered. This group forms the preregistered primary analytic sample ($n = 1,654$).

Analyses focused on the two preregistered primary outcomes, (i) support for the entire climate policy package and (ii) the incentive-compatible donation decision, and an exploratory outcome, (iii) support for each of the elements of the climate policy package. We estimated treatment effects on climate policy support using ordinary least squares (OLS) regression models with robust standard errors, including dummy variables for the two AI-generated narrative conditions and using the Control condition as the reference group. For the donation outcome, we estimated a linear probability model for consistency across outcomes. All analyses were conducted in Stata 19.

Exploratory analyses assessed whether psychological distance, future vividness, empathic engagement, future-oriented emotions, and perceived likelihood and valence of future scenarios were linked to treatment assignment and could help explain any differences observed across groups. These analyses were preregistered as research questions and conducted using structural equation modeling (SEM) in R. Since they were not preregistered as confirmatory mediation tests, we interpret these findings with caution.

As robustness checks, we also estimated models including preregistered sociodemographic covariates (age, gender, marital status, household income, education level, ethnicity, and political identity). Because only two primary outcomes were preregistered, we did not apply multiple comparison corrections to those analyses. Full model specifications, robustness checks, and ethical approval information are provided in the Appendix.

**2.6. Sensitivity analyses**

As a robustness check, we re-estimated all primary models, including preregistered demographic covariates (age, gender, marital status, household income, education, ethnicity, and political identity). Results were substantively unchanged. Full adjusted estimates and additional robustness checks are provided in Appendix C.

**3. Results**

**3.1. Pre-registered analysis**

**3.1.1. Effects on climate policy support and donation behavior**

We first examined whether AI-generated letters from the future increased parents' support for climate policies and willingness to donate to an environmental charity, and whether personalization enhanced these effects relative to a generic future-person narrative. Figure 2 presents unadjusted and covariate-adjusted treatment effects for both outcomes.

Across both outcomes, neither narrative message produced a statistically significant effect relative to the fact-based control. For example, for climate policy support, estimated effects were close to zero in both the Future Person condition ($b = –0.01$, 95% CI [–0.20, 0.18], $p = .93$) and the Future Child condition ($b = –0.05$, 95% CI [–0.25, 0.14], $p = .60$). Effects on donation behavior were similarly small and non-significant (Future Person: $b = 0.03$, 95% CI [–0.02, 0.09], $p = .24$; Future Child: $b = 0.04$, 95% CI [–0.02, 0.10], $p = .16$). Covariate adjustment did not meaningfully change these conclusions, and no treatment contrast reached statistical significance.

Consistent with these findings, there was little evidence that personalization improved message impact: The Future Child and Future Person conditions did not significantly differ on either outcome ($p = .68$ for policy support; $p = .82$ for donation behavior). Overall, these results do not support the preregistered hypotheses that AI-generated narrative messages would increase climate policy support or charitable giving, nor that personalization would strengthen these effects.

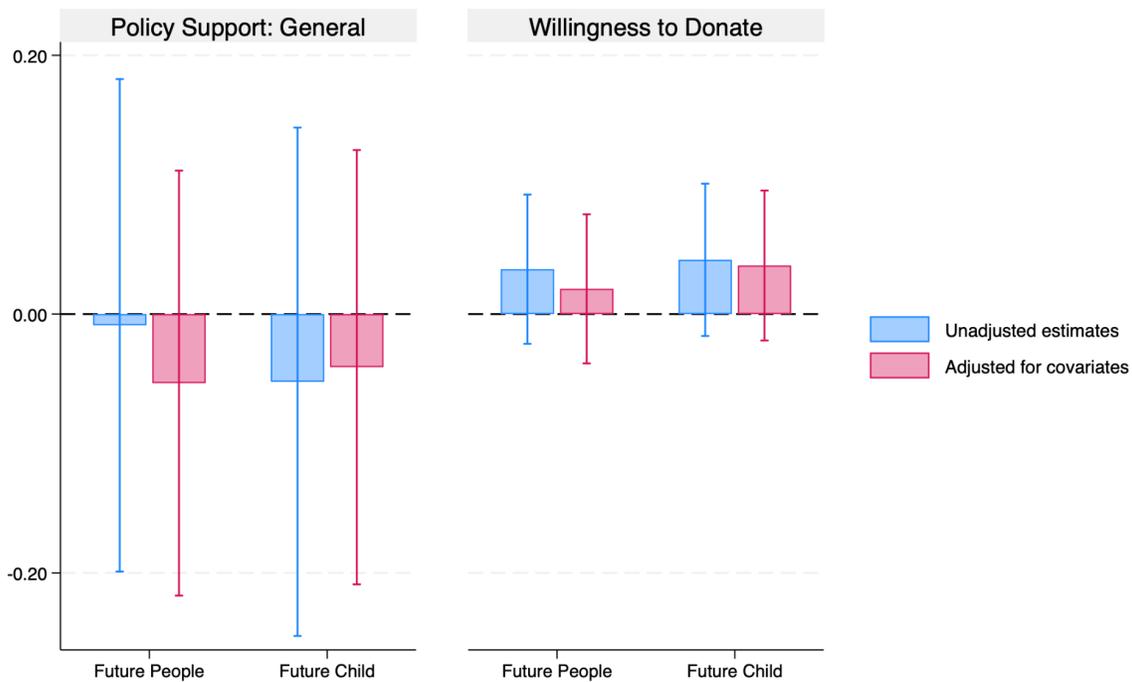

**Figure 2.** Effects of future-oriented messaging on climate policy support and willingness to donate compared to a fact-based climate report ($N = 1,654$). Bars represent OLS point estimates for two treatments—letters from a hypothetical *Future Person* or *Future Child*—on support for general climate policy (left panel) and willingness to donate a monetary bonus to a reforestation charity (right panel). Blue bars show unadjusted estimates; red bars reflect models adjusted for the order of scenario presentation, child's age, and participant demographics (age, gender, marital status, education, and political identity). Error bars indicate 95% confidence intervals. Policy support was measured on a 1 (strongly oppose) to 7 (strongly support) Likert scale; willingness to donate is a binary indicator.

As shown in Figure 3, we further examined whether either narrative messages influenced support for specific climate policies compared to the fact-based report. Our exploratory analysis showed that estimated treatment effects were small and statistically non-significant across both narrative messages and all eight policies, with confidence intervals consistently spanning zero. For example, in the model predicting support for heavily taxing coal, the estimated effects were 0.08 (95% CI: –0.15 to 0.29, $p = .50$) for the Future Person condition and 0.03 (95% CI: –0.20 to 0.26, $p = .79$) for the Future Child condition. Similarly, in the model predicting support for heavily subsidizing renewable energy, the effects were 0.02 (95% CI: –0.17 to 0.29, $p = .86$) and 0.10 (95% CI: –0.08 to 0.28, $p = .29$), respectively. While point estimates were sometimes directionally positive—for example, for subsidizing nuclear energy and reductions in military spending—the overall pattern showed no consistent evidence that either message increased support for individual policies compared to the fact-based control condition.

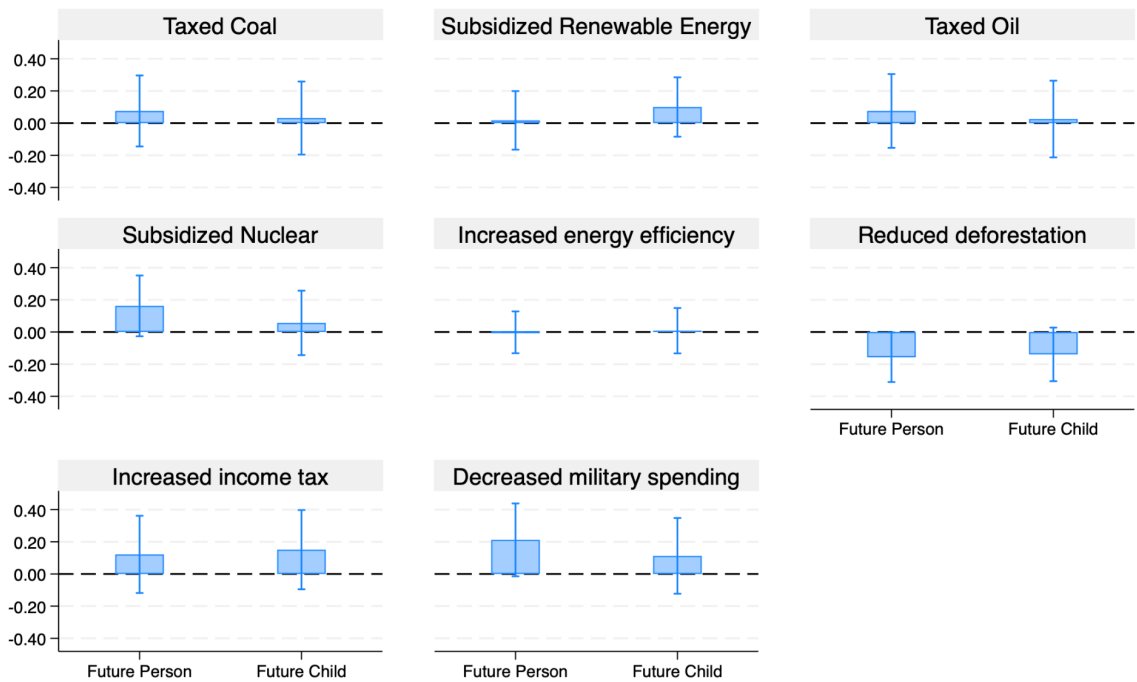

**Figure 3.** Unadjusted effects of future-oriented messaging on support for specific climate-related policies compared to a fact-based climate report ($n = 1,654$). Bars show unadjusted OLS point estimates for the effect of receiving a letter from *Future Person* or *Future Child* on support for eight individual climate-related policies. Each policy support measure was rated on a 1 (strongly oppose) to 7 (strongly support) Likert scale. Outcomes include support for: (1) highly taxing coal, (2) highly subsidizing renewable energy, (3) highly taxing oil, (4) highly subsidizing nuclear energy, (5) highly increasing energy efficiency, (6) highly reducing deforestation, (7) increasing income tax, and (8) decreasing military spending. Error bars represent 95% confidence intervals.

### 3.1.2. Psychological distance, vividness, empathy, and future-oriented emotions

We next explored participants' psychological responses to the narrative messages. These analyses were preregistered as exploratory and were designed to assess whether the messages influenced perceived psychological distance, future vividness, empathic engagement, and future-oriented emotions. This section examines how these constructs varied across conditions and whether they may help illuminate participants' engagement with the narratives.

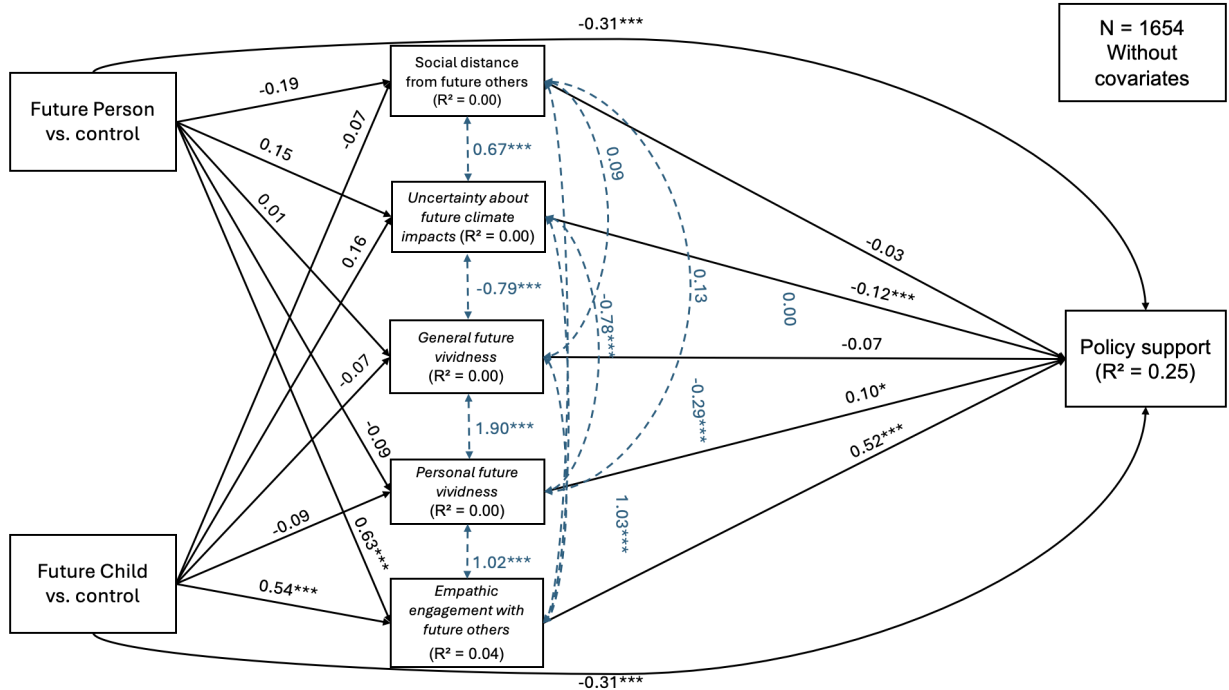

**Figure 4.** Path analysis of the unadjusted effects of future-oriented messaging on climate policy support via psychological distance, perceived vividness, and empathic engagement ($N = 1,654$). Structural equation model testing how *Future Person* and *Future Child* framings (vs. control) influence policy support through five psychological distance mediators: social distance from future others, uncertainty about future climate impacts, general future vividness, personal future vividness, and empathic engagement with future others. Solid lines indicate direct effects; dashed lines represent correlations among mediators. * $p < .05$, ** $p < .01$, *** $p < .001$. The model explains 25% of the variance in policy support ($R^2 = .25$).

Figure 4 summarizes exploratory path models examining associations between message condition, psychological responses, and policy support. Compared to the fact-based control message, both narrative conditions were associated with substantially higher empathic engagement with individuals living in 2050 (Future Person: $b = 0.63$, 95% CI [0.47, 0.79], $p < .001$; Future Child: b = 0.54, 95% CI [0.37, 0.71], p < .001). Empathic engagement was, in turn, positively associated with climate policy support ($b = 0.52$, 95% CI [0.45, 0.60], $p < .001$).

By contrast, there was limited evidence that either message meaningfully altered perceived social and hypothetical psychological distance. Estimated effects for uncertainty about future climate impacts were small and not statistically significant when comparing the narrative conditions to the control (Future Person: $b = 0.15$, 95% CI [–0.06, 0.36], $p = .15$; Future Child: $b = 0.16$, 95% CI [–0.05, 0.37], $p = .13$). Likewise, neither message showed clear effects on future vividness (e.g., Future Child on personal vividness: $b = –0.09$, 95% CI [–0.28, 0.09], $p = .32$).

In addition, when psychological variables were entered into the model simultaneously, both narrative conditions showed significant negative direct associations with policy support relative to the control group (Future Person: $b = –0.31$, 95% CI [–0.48, –0.15], $p < .001$; Future Child: $b = –0.31$, 95% CI [–0.48, –0.14], $p < .001$). This pattern suggests that the overall null effects observed in the preregistered analyses may reflect offsetting associations at the psychological level, whereby increases in empathic engagement coincide with other processes associated with lower policy support. Because the associations between mediators and the outcome are observational rather than experimentally manipulated, we interpret these mediation patterns cautiously and explore them further in the subsequent section.

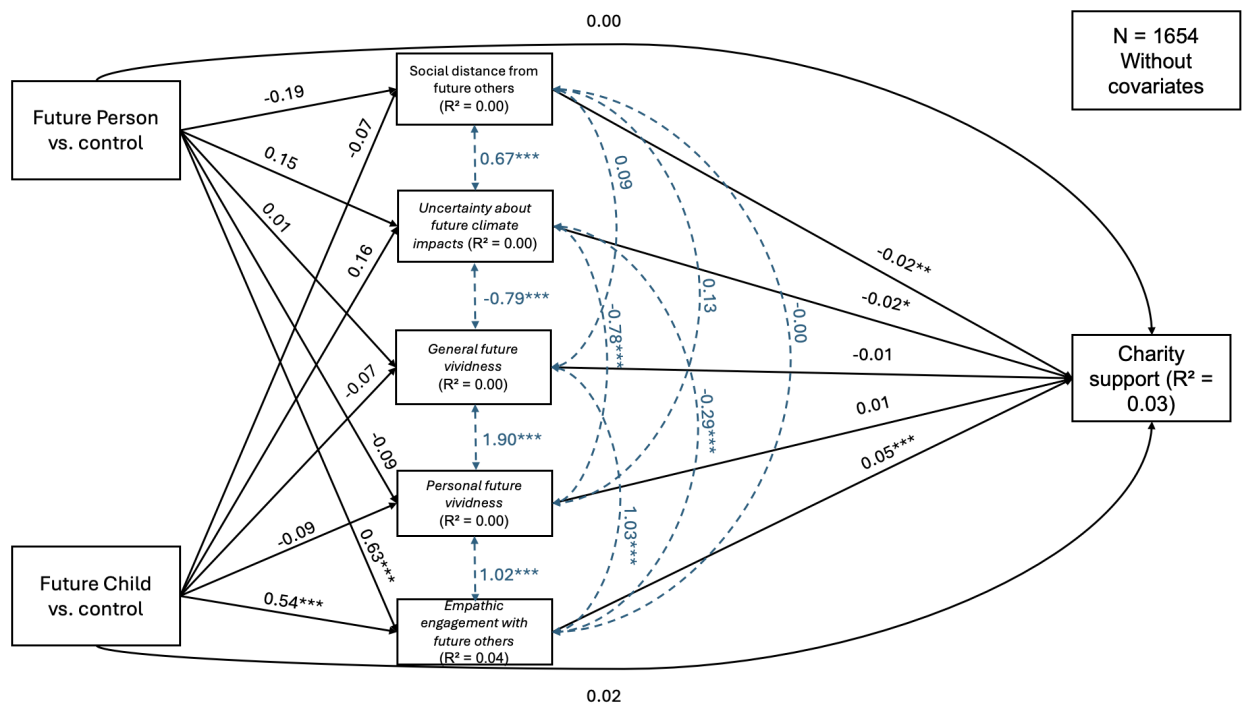

**Figure 5.** Path analysis of the unadjusted effects of future-oriented messaging on donation to *One Tree Planted* via psychological distance, perceived vividness, and empathic engagement ($n = 1,654$). Structural equation model testing how *Future Person* and *Future Child* framings (vs. control) influence donation through five psychological distance mediators: social distance from future others, uncertainty about future climate impacts, general future vividness, personal future vividness, and empathic engagement with future others. Solid lines indicate direct effects; dashed lines represent correlations among mediators. * $p < .05$, ** $p < .01$, *** $p < .001$. The model explains 3% of the variance in donation ($R^2 = .03$).

Patterns for charitable donation were broadly similar. Both narrative conditions were associated with greater empathic engagement than the fact-based control, and this emotional connection was positively associated with willingness to donate to the reforestation charity. Each one-point increase in empathic engagement was associated with approximately a five-percentage-point increase in donation likelihood ($b = 0.05$, 95% CI [0.03, 0.07], $p < .001$).

Greater perceived social distance from future others and greater uncertainty about future climate impacts were both negatively associated with donation likelihood (social distance: $b = –0.02$, 95% CI [–0.03, –0.01], $p < .001$; uncertainty: $b = –0.01$, 95% CI [–0.03, –0.001], $p = .03$). However, neither message substantially shifted these constructs relative to the control condition.

Direct associations between the messages and donation behavior remained small and statistically non-significant in the exploratory structural models (Future Person: $b = 0.00$, 95% CI [–0.05, 0.06], $p = .99$; Future Child: $b = 0.02$, 95% CI [–0.04, 0.07], $p = .52$). This pattern suggests that any associations between narrative conditions and donation behavior in these exploratory models were primarily accounted for by individual differences in psychological responses rather than by direct effects of message condition. Full results are reported in Supplementary Table S4.

Figures 6 and 7 examine whether narrative messages were associated with shifts in future-oriented emotional responses—namely, relaxation, hope, motivation to act, guilt, and fear—that could plausibly be linked to climate policy support and donation behavior. As shown in Figure 6, motivation and fear were the two emotional responses most strongly associated with support for climate policy. A one-point increase in self-reported motivation to act corresponded to a 0.57-point increase in policy support (95% CI [0.51, 0.63], $p < .001$), and a one-point increase in fear was associated with a 0.25-point increase (95% CI [0.19, 0.32], $p < .001$).

However, neither narrative message showed consistent associations with these emotional responses. Across conditions, there were no statistically significant differences in motivation, fear, hope, or guilt. The one exception was relaxation: the Future Child message was associated with slightly lower feelings of relaxation about the future ($b = –0.14$, 95% CI [–0.28, –0.002], $p = .046$). Lower relaxation, in turn, was negatively associated with policy support ($b = –0.29$, 95% CI [–0.36, –0.22], $p < .001$). A comparable pattern was observed for donation behavior, as shown in Figure 7.

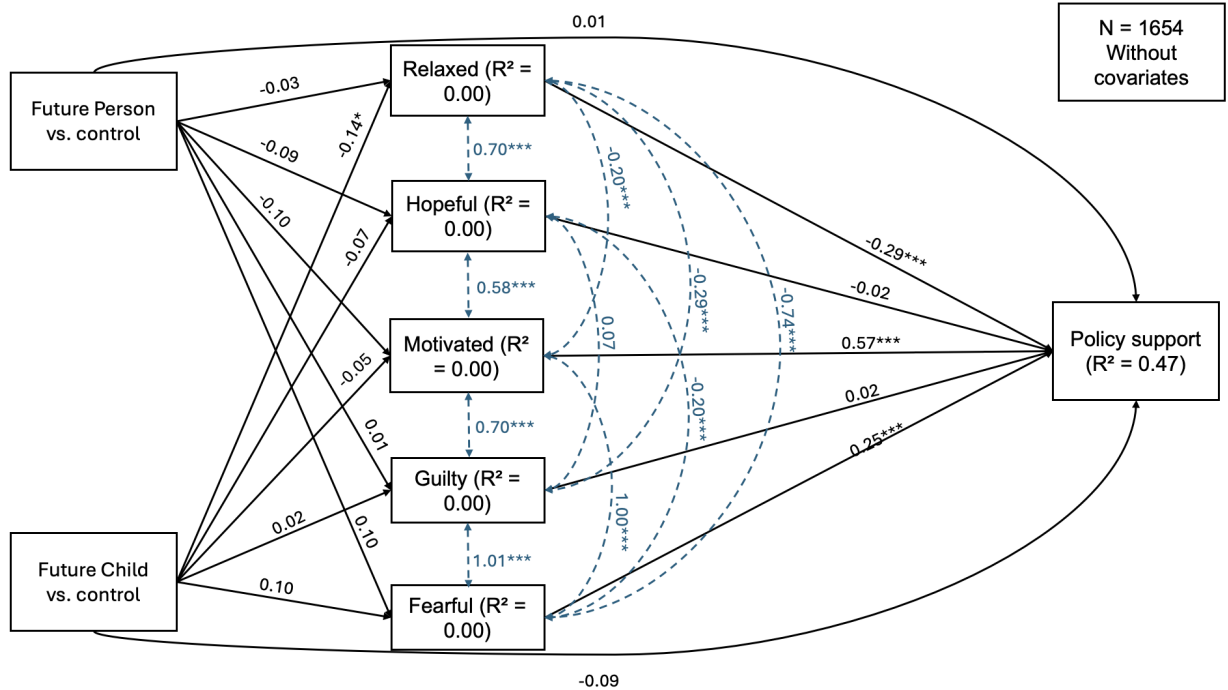

**Figure 6.** Path analysis of the unadjusted effects of future-oriented messaging on climate policy support via future-oriented emotions ($N$ = 1,654). Structural equation model testing how *Future Person* and *Future Child* framings (vs. control) influence policy support through five future-oriented emotions mediators: relaxed, hopeful, motivated to act, guilt, and fear. Solid lines indicate direct effects; dashed lines represent correlations among mediators. * $p < .05$, ** $p < .01$, *** $p < .001$. The model explains 47% of the variance in policy support ($R^2 = .47$), nearly twice as much as psychological distance, perceived vividness, and empathic engagement ($R^2 = .25$), suggesting that emotions are more immediately connected to policy support compared to other mechanisms.

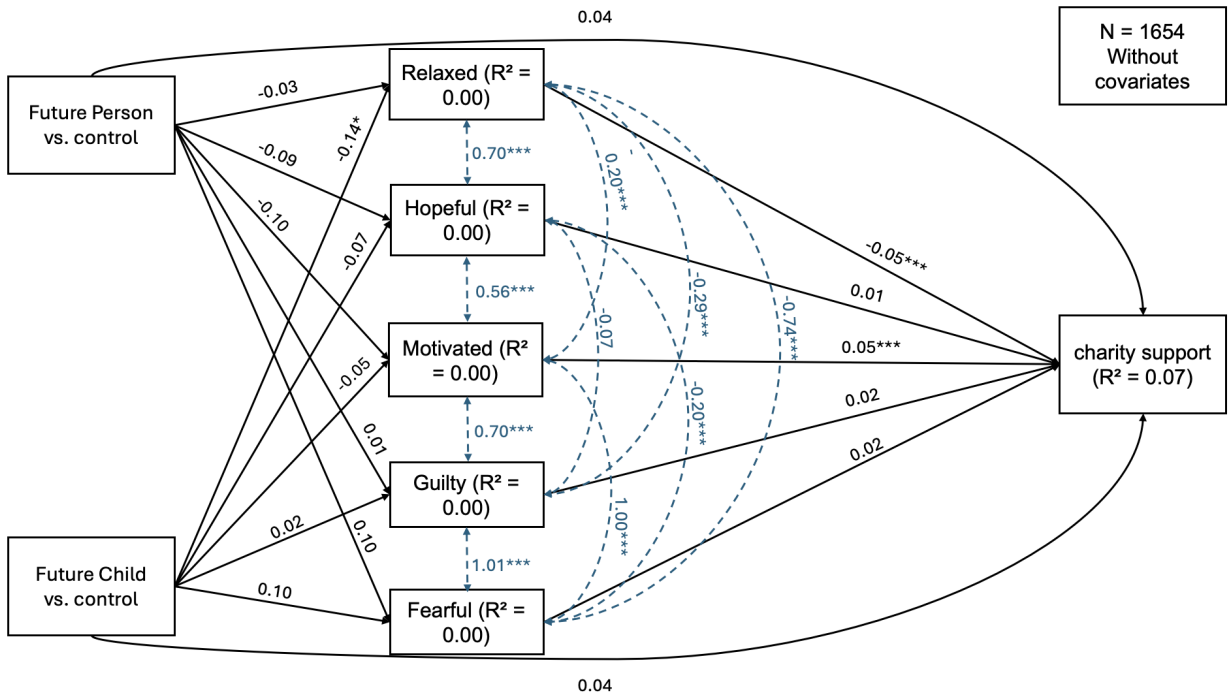

**Figure 7.** Path analysis of the unadjusted effects of future-oriented messaging on donation to *One Tree Planted* via future-oriented emotions (N=1,654). Structural equation model testing how *Future Person* and *Future Child* framings (vs. control) influence donation through five future-oriented emotions mediators: relaxed, hopeful, motivated to act, guilt, and fear. Solid lines indicate direct effects; dashed lines represent correlations among mediators. * $p < .05$, ** $p < .01$, *** $p < .001$. The model explains 7% of the variance in donation ($R^2 = .07$).

### 3.1.3. Summary of the exploratory findings

Taken together, the exploratory analyses suggest that the narrative messages elicited increased empathic engagement with future generations, and that empathy was positively associated with both policy support and charitable giving. However, the magnitude of these associations was insufficient to produce detectable changes in either outcome in the preregistered analyses, and the narratives showed only limited associations with other psychological constructs identified in prior research. Measures of psychological distance and future vividness showed little variation across conditions, and emotional predictors, such as fear and the motivation to act, remained largely unchanged.

These patterns indicate that participants engaged emotionally with the narratives, but this engagement did not translate into broader shifts in the psychological processes most strongly linked to pro-environmental action. We therefore interpret these findings descriptively and without causal inference.

In the next section, which was not preregistered, we conduct follow-up analyses examining how participants evaluated the two envisaged futures, including the perceived likelihood and desirability of each scenario, as a possible factor underlying the null treatment effects.

### 3.2. Exploratory, non-preregistered analysis

To further contextualize the preregistered results, we conducted exploratory analyses of how participants appraised the two contrasting future scenarios presented in the narrative conditions—one in which society acts to mitigate climate change, and one in which it does not. Whereas the preceding analyses examined psychological processes in aggregate, this section focuses specifically on participants' emotional and cognitive responses to each scenario separately. For clarity, we limit the primary analysis to policy support (see Figure 8).

Both narrative conditions were associated with more negative emotional appraisals of the undesirable future (no-action scenario) and more positive appraisals of the desirable future (action scenario) relative to the control group. This pattern suggests greater contrast in how participants evaluated the two scenarios after reading the narrative messages.

At the same time, the narrative conditions were associated with a lower perceived likelihood that the desirable future would occur. The Future Person condition was associated with a 0.42-point reduction in the perceived likelihood of the desirable future (95% CI [–0.62, –0.21], $p < .001$), and the Future Child condition with a 0.30-point reduction (95% CI [–0.51, –0.09], $p = .005$). Estimates for the likelihood of the undesirable future were comparatively small and not statistically significant.

Although interpretation should be cautious given the exploratory nature of these analyses, these patterns suggest that while the narrative messages were associated with more emotionally divergent appraisals of the two futures, they may also have been associated with greater perceived difficulty in achieving the favorable outcome.

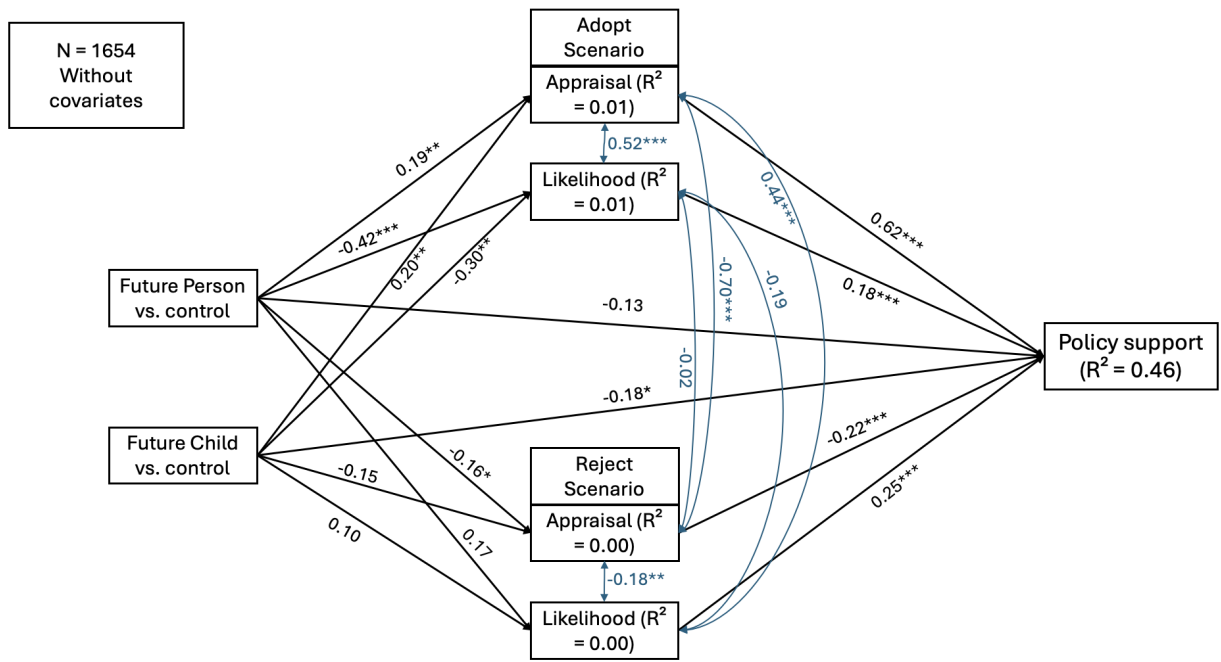

**Figure 8.** Path analysis of the unadjusted effects of future-oriented messages on climate policy support via appraisals of two distinct future scenarios ($N = 1,654$). Structural equation model testing how *Future Person* and *Future Child* framings (vs. control) influence support for climate policy through emotional and cognitive responses to two imagined futures: one in which society fails to act on climate change (Reject Scenario), and one in which collective action succeeds (Adopt Scenario). Mediators include emotional appraisal and perceived likelihood for each scenario. Solid lines indicate direct effects; curved lines indicate correlations among mediators. $p < .05$, $p < .01$, $p < .001$. Model explains 46% of the variance in policy support ($R^2 = .46$).

Both emotional appraisals and perceived likelihoods of the two future scenarios were strongly associated with climate policy support. A more positive appraisal of the desirable future was positively associated with policy support ($b = 0.62$, 95% CI [0.53, 0.70], $p < .001$), whereas a more positive appraisal of the undesirable future was negatively associated ($b = -0.22$, 95% CI [-0.29, -0.15], $p < .001$). Perceived likelihood also mattered: higher perceived likelihood of both the desirable ($b = 0.25$, 95% CI [0.21, 0.29], $p < .001$) and undesirable ($b = 0.18$, 95% CI [0.14, 0.22], $p < .001$) futures was associated with greater support for climate action.

In these exploratory models, the Future Child message showed a significant negative direct association with policy support ($b = -0.18$, 95% CI [-0.32, -0.03], $p = .02$), while the direct association for the Future Person message was smaller and not statistically significant ($b = -0.13$, 95% CI [-0.27, 0.01], $p = .07$). Consistent with the earlier exploratory findings, these results suggest that psychological responses to the narrative messages were complex and multifaceted, and should be interpreted cautiously.

Together, these exploratory analyses indicate that although the narrative messages were associated with more differentiated emotional and cognitive appraisals of possible climate futures, they were also associated with lower perceived likelihood of the desirable future. This pattern may help contextualize the absence of detectable effects on the preregistered behavioral outcomes.

## 4. Discussion

This study examined whether AI-generated narrative messages, presented as letters from the future, could increase engagement with climate action among parents in the United States. Although both narrative conditions increased empathic engagement with people living in 2050, neither produced detectable changes in climate policy support or related charitable giving in the preregistered analyses. Furthermore, personalizing the narrative as originating from a participant's future child did not meaningfully enhance its impact relative to a message from a generic future person. These findings suggest that, while generative AI can elicit emotional responses in climate communication, emotional vividness and kinship-based personalization were insufficient to increase behavioral engagement in this context. Understanding why heightened emotional engagement did not translate into behavioral change is therefore central to interpreting these findings.

Although null results are often viewed as inconclusive, in this case, they provide valuable insights. Generative AI is rapidly becoming a preferred tool for large-scale, personalized persuasion, yet there is limited empirical evidence on whether such emotionally intense, future-focused narratives actually influence behavior. Using a preregistered, well-powered experiment, our results highlight clear limits to what AI-driven personalization can achieve, even when emotional engagement increases. Clarifying when these interventions fail is essential for guiding responsible deployment and for tempering overly optimistic expectations about AI-mediated public communication. This helps establish boundary conditions for emotionally-charged, technology-based interventions.

Beyond behavioral outcomes, the findings offer insight into how individuals cognitively and emotionally engage with AI-generated persuasive content. The exploratory analyses suggest that participants formed more differentiated emotional appraisals of desirable and undesirable futures, yet simultaneously perceived the favorable outcome to be less likely. While interpretation is necessarily cautious, this pattern may indicate that when AI systems present multiple contrasting outcomes within a single message, they may inadvertently undermine the perceived feasibility of positive futures. This is consistent with evidence suggesting that persuasive impact depends not only on emotional resonance but also on perceived credibility, coherence, and authenticity (Moser, 2010; Sanz-Menéndez & Castro, 2019).

These findings also contribute to ongoing discussions about psychological distance and environmental engagement. Although empathic concern increased, other dimensions of psychological distance such as perceived uncertainty and personal connectedness did not change. This suggests that reducing psychological distance may require more than enhancing emotional vividness, particularly when messages are generated algorithmically. Single exposures may also be insufficient to shift deeper beliefs about climate futures or personal relevance. More broadly, these results reinforce the need to understand which dimensions of psychological distance are most malleable, and how emotional tone, framing, and perceived plausibility interact to shape engagement.

These results highlight important design considerations for AI-based communication tools. Emotional personalization alone appears insufficient to increase climate engagement, suggesting that future systems should foreground not only emotional intensity but also credibility cues, coherence, and authenticity. Designers may need to test whether single-outcome narratives, or interactive formats that allow users to explore scenarios sequentially rather than simultaneously, better sustain perceived plausibility. Incorporating user control (e.g., selecting narrative tone, timescale, or policy pathways) may also improve perceived authenticity and reduce skepticism. Finally, systematic pretesting of AI-generated messages—especially for unintended impressions such as artificiality, exaggeration, or implausibility—will be essential to ensure that the scalability benefits of generative AI do not come at the cost of diminished trust or behavioral engagement.

Several limitations warrant acknowledgment. First, the sample was recruited online and may not reflect the broader population of parents with respect to climate attitudes or familiarity with digital media. Second, the intervention consisted of a single exposure to two different scenarios; repeated or longitudinal interaction with future-oriented narratives may yield different effects. Third, although we examined a range of psychological responses, we did not directly measure perceived authenticity, trust, or discomfort, which may be central to understanding audience reactions to AI-generated content. Fourth, we tested only one set of climate scenarios and one implementation of AI-generated narratives. Future research should examine alternative narrative formats, modes of interaction, emotional tones, and model designs. Finally, caution is warranted in generalizing these findings beyond climate communication.

In sum, while AI-generated, emotionally personalized narratives can increase empathic engagement, this study finds no consistent evidence that they translate into behavioral change. The results highlight the need to balance emotional resonance with trust, credibility, and narrative plausibility when designing technologically mediated interventions, and point to important constraints that must be understood for such tools to be effective in practice.

**Data Availability Statement**

All materials, including the preregistration document, experimental stimuli, raw data, and analysis code, are openly available at the Open Science Framework: [https://osf.io/uq2w9/?view_only=c9fff96f26544c5d8f1c874a9b9b41ee](https://osf.io/uq2w9/?view_only=c9fff96f26544c5d8f1c874a9b9b41ee).

**Declaration of Generative AI and AI-Assisted Technologies in the Manuscript Preparation Process**

During the preparation of this work, the authors used generative AI in two capacities. First, GPT-4 was used to generate the experimental stimuli (personalized climate narratives from future persons) that were presented to participants as part of the research design, as described in detail in the Methods section. Second, the authors used Claude (Anthropic) and Grammarly to improve language clarity, grammar, and readability during manuscript preparation. After using these tools, the authors carefully reviewed, edited, and revised all content to ensure accuracy and that the manuscript represents the authors' original intellectual contribution, analysis, and interpretation. The authors take full responsibility for the content of this published article.

# References

Allcott, H., & Rogers, T. (2014). The short-run and long-run effects of behavioral interventions: Experimental evidence from energy conservation. *American Economic Review*, *104*(10), 3003–3037.

Brügger, A., Morton, T. A., & Dessai, S. (2016). "Proximising" climate change reconsidered: A construal-level theory perspective. *Journal of Environmental Psychology, 46*, 125–135.

Carlsson, F., Gravert, C., Johansson-Stenman, O., & Kurz, V. (2021). The use of green nudges as an environmental policy instrument. *Review of Environmental Economics and Policy*, *15*(2), 216-237.

Chatterjee, J. S., Thongprasert, S., & Some, S. (2024). Communication, Climate Mitigation, and Behavior Change Interventions: Understanding Message Design and Digital Media Technologies. *Annual Review of Environment and Resources*, *49*, 655–672.

Fornwagner, H., & Hauser, O. P. (2022). Climate action for (my) children. *Environmental and Resource Economics*, *81*(1), 95–130.

Gifford, R. (2011). The dragons of inaction: psychological barriers that limit climate change mitigation and adaptation. *American Psychologist*, *66*(4), 290–302.

Hershfield, H. E., Cohen, T. R., & Thompson, L. (2012). Short horizons and tempting situations: Lack of continuity to our future selves leads to unethical decision making and behavior. *Organizational Behavior and Human Decision Processes*, *117*(2), 298–310.

Jones, C. R., Hine, D. W., & Marks, A. D. G. (2017). The future is now: Reducing psychological distance to increase public engagement with climate change. *Risk Analysis, 37*(2), 331–341. https://doi.org/10.1111/risa.12601.

Keller, A., Marsh, J. E., Richardson, B. H., & Ball, L. J. (2022). A systematic review of the psychological distance of climate change: Towards an evidence-based construct. *Journal of Environmental Psychology, 81*, Article 101822. https://doi.org/10.1016/j.jenvp.2022.101822.

Lawson, D. F., Stevenson, K. T., Peterson, M. N., Carrier, S. J., Strnad, R. L., & Seekamp, E. (2019). Children can foster climate change concern among their parents. *Nature Climate Change, 9*, 458–462.

Liberman, N., & Trope, Y. (2008). The psychology of transcending the here and now. *Science*, *322*(5905), 1201-1205.

Loewenstein, G., & Chater, N. (2017). Putting nudges in perspective. *Behavioural Public Policy*, *1*(1), 26-53.

McDonald, R. I., Chai, H. Y., & Newell, B. R. (2015). Personal experience and the 'psychological distance' of climate change: An integrative review. *Journal of Environmental Psychology*, *44*, 109–118.

Milfont, T. L. (2010). Global warming, climate change and human psychology. *Psychological approaches to sustainability: Current trends in theory, research and practice*, *19*, 42.

Moser, S. C. (2010). Communicating climate change: history, challenges, process and future directions. Wiley Interdisciplinary Reviews: Climate Change, 1(1), 31-53.

Norgaard, K. M. (2011). *Living in denial: Climate change, emotions, and everyday life*. MIT Press.

OECD. (2024). *The Climate Action Monitor 2024: Monitoring progress towards net zero* (IPAC Synthesis Report). OECD Publishing.

Pahl, S., & Bauer, J. (2013). Overcoming the distance: Perspective taking with future humans improves environmental engagement. *Environment and Behavior*, *45*(2), 155-169.

Pataranutaporn, P., Doudkin, A., & Maes, P. (2025). OceanChat: The effect of virtual conversational AI agents on sustainable attitude and behavior change. *arXiv preprint arXiv:2502.02863*.

Pataranutaporn, P., Winson, K., Yin, P., Lapapirojn, A., Ouppaphan, P., Lertsutthiwong, M., Maes, P., & Hershfield, H. E. (2024, October). Future you: a conversation with an AI-generated future self reduces anxiety, negative emotions, and increases future self-continuity. In *2024 IEEE Frontiers in Education Conference (FIE)* (pp. 1–10). IEEE.

Pi, Y., Pan, X., Slater, M., & Świdrak, J. (2025). Embodied time travel in VR: from witnessing climate change to action for prevention. *Frontiers in Virtual Reality*, *5*, 1499835.

Sanz-Menéndez, L., & Cruz-Castro, L. (2019). The credibility of scientific communication sources regarding climate change: A population-based survey experiment. Public Understanding of Science, 28(5), 534-553

Slovic, P. (2007). "If I look at the mass I will never act": Psychic numbing and genocide. *Judgment and Decision Making*, *2*(2), 79-95.

Spence, A., & Pidgeon, N. (2010). Framing and communicating climate change: The effects of distance and outcome frame manipulations. *Global Environmental Change*, *20*(4), 656–667.

Spence, A., Poortinga, W., & Pidgeon, N. (2012). The psychological distance of climate change. *Risk Analysis, 32*(6), 957–972.

Syropoulos, S., Markowitz, E. M., Demarest, B., & Shrum, T. (2023). A letter to future generations: Examining the effectiveness of an intergenerational framing intervention. *Journal of Environmental Psychology*, *90*, 102074.

Thoma, S. P., Hartmann, M., Christen, J., Mayer, B., Mast, F. W., & Weibel, D. (2023). Increasing awareness of climate change with immersive virtual reality. *Frontiers in Virtual Reality*, *4*, 897034.

Trope, Y., & Liberman, N. (2010). Construal-level theory of psychological distance. *Psychological review*, *117*(2), 440-463.

Vlasceanu, M., Doell, K. C., Bak-Coleman, J. B., & the Intervention Tournament Team. (2024). Addressing climate change with behavioral science: A global intervention tournament in 63 countries. *Science Advances, 10*(6), eadj5778.

Zhao, J., & Chen, F. S. (2023). i-Frame interventions enhance s-frame interventions. *Behavioral and Brain Sciences*, *46*, e180.

**Appendix A: Additional details of the procedures**

**Ethics**

We have secured ethical approval from the Institutional Review Board at one of the author's universities (Exempt ID: E-6515). All participants provided informed consent before beginning the survey experiment. Participants received a US$2 participation fee and an additional US$0.50 bonus, which they could either keep or donate to the *One Tree Planted* charity.

*Checks*

To ensure high data quality, we implement a bot, a comprehension, and an attention check.

For the bot check, we will use Qualtrics' invisible reCAPTCHA to detect potential bots. As part of the comprehension check, participants were asked to identify the year referenced in the scenarios by selecting from the given options: 2030, 2050, or 2100. This ensures their understanding of the report they have just read. In addition, participants completed two attention check questions to ensure they were actively engaged in the survey. They were asked to select a specific color from a list, with clear instructions indicating that they must choose "Green", and they were asked to indicate their agreement with the statement, "I ride my bike from California to New York to get to work every day." Responses will be recorded on a five-point scale ranging from 1 *Strongly disagree* to 5 *Strongly agree*, with 4 *Agree* and 5 *Strongly agree* being marked as inattentive.

*Demographics*

Participants responded to questions about their child's age, their age, gender, marital status, annual household income before tax, education level, ethnicity, and political identity.

**Procedure**

We describe the procedure of the between-person experiment, divided into three sections: pre-intervention, intervention, and post-intervention. The mean completion time was 14.20 minutes (*SD=7.67 minutes)*.

*Pre-intervention*

At the beginning of the experiment, all participants will provide informed consent. During this process, the study will be presented as 'Perspectives on Future Scenarios' to minimize self-selection bias and prevent inadvertently influencing participants' responses. After obtaining informed consent, participants will complete an attention check to ensure automated bots do not generate responses. Following this, they will be asked to provide demographic information, including their child's age, their own age, gender, marital status, annual household income before tax, education level, ethnicity, and political identity.

*Intervention*

Once participants complete the demographic questionnaire, they will be randomly assigned to one of three experimental conditions: the Control Group (Active Control), the Future Person Narrative Group, or the Future Child Narrative Group. Each condition consists of a positive and negative future scenario, and participants can choose which scenarios to view in what order.

*Post-intervention*

After a brief comprehension check—asking participants to identify the year in which the messages were written—we will assess their perceptions of the two future scenarios. Specifically, participants will report how likely they believe each scenario is to occur, their emotions associated with both futures, their perceived psychological distance from the people in these scenarios, their emotions related to climate change, and their level of support (in fixed order). Additionally, those in the Future Child Narrative group will evaluate the realism of the letter from the future. Participants will then complete a final attention check before indicating their preferred allocation of a $0.50 bonus. They will choose whether to donate the full $0.50 to One Tree Planted, or keep the entire $0.50 for themselves. The survey ends with a thank you note and a debriefing.

**AI-generated messages used in each treatment**

**1) Control treatment**

**Positive scenario: the public majority supports the climate policy**

Status Report from 2050

Energy & Infrastructure:

- Solar panels and vertical gardens are integrated into city architecture
- Electric cars are powered by building-integrated solar panels
- Energy costs are minimal due to renewable energy subsidies
- Smart grid technology manages power distribution efficiently
- Buildings maintain comfort year-round with minimal energy use

Environmental Impact:

- Global temperatures have stabilized
- Nature shows signs of recovery
- Previously barren areas are now thriving forests

- Wildlife populations have increased in reforested areas
- Carbon emissions are significantly reduced

Economic Changes:

- Taxes need to increase to successfully fund green initiatives
- Military budgets need to be reduced to support reforestation
- Renewable energy costs are lower than those in the fossil fuel era
- Local vertical farming reduces transportation costs
- Clean technology innovation exceeded expectations

Quality of Life:

- Year-round access to locally grown fresh produce
- Improved air quality in urban areas
- More green spaces for community gatherings
- Efficient public transportation systems
- Stable climate conditions for outdoor activities

**Negative scenario: the public majority rejects the climate policy**

Status Report from 2050

Energy & Infrastructure:

- Cities are frequently covered in smog
- Coal plants still operating with "clean coal" permits
- Power grid fails during extreme weather
- Electric vehicle infrastructure has never developed
- Buildings lack energy efficiency upgrades

Environmental Impact:

- Global temperatures exceed worst projections
- Widespread deforestation
- Frequent wildfires destroy the remaining forests
- Coastal cities threatened by rising seas
- Regular air quality alerts restrict outdoor activities

Economic Changes:

- Fossil fuel industry maintains dominance

- Military spending increased for climate crisis response
- High food prices due to supply chain disruptions
- Insurance companies abandon flood-prone areas
- Carbon capture technology remains unrealized

Quality of Life:

- Extended summer heat waves
- Limited access to fresh produce
- Reduced outdoor community spaces
- Wealth gap in climate protection
- Regular power outages during extreme weather

**2) "Future Person" treatment**

**Positive scenario: the public majority supports the climate policy**

Dear Reader,

I'm writing to you from 2050. You wouldn't believe how different things are here in 2050! Twenty years ago, at the end of 2030, governments around the world finally came together to implement important climate policies. Fortunately, these policies seemed to have largely worked and the worst predictions were avoided. In some places, things are even better than in 2025!

From my apartment window, for instance, I can see a skyline dotted with vertical gardens and solar panels that seamlessly blend with the architecture.

My daily life is so different from what you know. Most people now use solar power from their building's integrated panels to charge their electric cars (thanks to those energy efficiency mandates!). The best part? It costs next to nothing to install and run, thanks to those subsidies. The income tax increases that everyone worried about back then? It turned out to be such a small price to pay for the quality of life we enjoy now.

Remember those endless debates about military spending? In 2030, countries finally agreed to reduce military spending to fund massive reforestation projects. You should see the forests now! Areas that were once barren are recovering and biodiversity is bouncing back. Last weekend, I went hiking in what used to be a dustbowl but the place is now teeming with butterflies and you can hear all sorts of birds singing. The preservation policies really worked.

All buildings must now be designed to meet energy efficiency standards - this made such a difference, too. My apartment stays comfortable year-round with minimal energy use. The smart grid technology we use now would seem like science fiction to you! And get this – many of our fresh fruits and vegetables are grown in local vertical farms. This reduces transportation emissions and provides us with fresh produce for much of the year.

It wasn't always easy. The transition had its challenges, but looking back, those policy decisions were crucial. The carbon taxes on fossil fuels pushed innovation in clean technology faster than anyone expected. Now, renewable energy is so cheap and efficient that it seems bizarre we ever relied on fossil fuels.

The best part? We're actually on track to meet our climate goals. Global temperatures have stabilized, and we're seeing the first signs of nature recovering. It's not perfect – we still have work to do – but we've proven that decisive action can make a difference.

I wish you could see this future. It's the one we fought for, and it's so much better than what could have happened if we had done nothing. Those policies that seemed so ambitious in 2025? They were just what we needed. We didn't just survive the climate crisis – in many ways, we created a better world.

Best wishes from 2050,

A Voice from the Future

P.S. Those reforestation projects have created the perfect spot for community gatherings - under a canopy of trees that weren't even saplings in your time!

**Negative scenario: the public majority rejects the climate policy**

Dear Reader,

I'm writing to you from 2050, and I wish I had better news about how things turned out after 2025. Remember those climate policies proposed back in 2025? The ones that could have changed everything? Well, the resistance was stronger than anyone anticipated.

From my apartment window, I see a city shrouded in smog most days. The proposed taxes on coal and oil were heavily watered down after massive industry pushback. The fossil fuel companies launched incredibly effective misinformation campaigns, convincing people that the tax increase would "destroy the economy." Instead of being phased out, coal plants were given extensions and "clean coal" permits. The renewable and nuclear energy subsidies were cut to a fraction of what was needed.

My daily life has become a struggle. Most people still drive gas-powered cars because the electric vehicle infrastructure never materialized and property developers blocked the energy efficiency mandates, claiming they were "too costly". My own apartment is barely habitable during the summer heat waves, which now last months. The promised smart grid remained a pipe dream, and our outdated power system frequently fails during extreme weather events.

The military budget cuts nations agreed upon in 2030 were reversed within years, with politicians citing "national security concerns" over climate refugees and resource conflicts. The money that could have gone to reforestation instead went to building walls on country borders and funding climate disaster response teams. You should see what's left of the forests. The rapid deforestation for "economic recovery" has left vast stretches of land barren. Last weekend, I visited the park where we used to have community gatherings – it's now a dustbowl, the trees lost to wildfires and drought.

Food prices are sky-high because we never developed those local vertical farms. Most fresh fruit and vegetables come from industrial greenhouses hundreds of miles away, when the extreme weather allows for growing at all. The supply chain disruptions we see now make those early COVID-19 pandemic shortages you lived through look trivial.

Back in 2025, many people argued that these policies were "too ambitious"? Now we know they weren't ambitious enough, and we didn't fight hard enough to protect them. The fossil fuel companies still run things. They kept promising that carbon capture technology would save us, but it was always 10 years away, then 15, and now... well.

Global temperatures have far exceeded our worst projections. Many coastal cities are fighting a losing battle against rising seas, and the insurance companies have abandoned most flood-prone areas. The wealthy live in climate-controlled compounds, while the rest of us endure the heat, storms, and bad air.

I wish I could show you a different future. One where we stood firm against the opposition to those policies. One where we didn't let short-term profits override our planet's future. The saddest part? We knew exactly what needed to be done in 2025. We had the technology, the policies, even the public support – but we let the momentum slip away.

With deep concern from 2050,

A Voice from the Future

P.S. Our community gatherings must now be held indoors due to the air quality alerts. The forest venues we once enjoyed don't exist anymore – they were lost in the great fires of 2039.

### 3) "Future Child" treatment

**Positive scenario: the public majority supports the climate policy**

Dear dad,

I hope this letter finds you well back in 2025! It's me, Pat, your child writing to you from Boston. I am now 35. You wouldn't believe how different things are here in 2050!

Twenty years ago, at the end of 2030, governments around the world finally came together to implement a comprehensive climate policy. Fortunately, this policy package seemed to have largely worked and the worst predictions avoided. In some places things are even better than in 2025! Our cities are absolutely beautiful now. From my apartment windowI can see a skyline dotted with vertical gardens and solar panels that seamlessly blend with the architecture.

My daily life is so different from what you know. I charge my electric car using solar power from our building's integrated panels (thanks to those energy efficiency mandates!). The best part? It costs me next to nothing to run, thanks to all those renewable energy subsidies. The income tax increase everyone worried about back then? It turned out to be such a small price to pay for the quality of life we enjoy now.

Remember those endless debates about military spending? In 2030, countries finally agreed to reduce military spending to fund massive reforestation projects. You should see the forests now! Areas that were once barren are recovering and biodiversity is bouncing back. Last weekend, I went hiking in what used to be a dustbowl but the place is now teeming with butterflies and you can hear all sorts of birds singing. The preservation policies really worked.

All buildings must now be designed to meet energy efficiency standards - this made such a difference, too! My apartment stays comfortable year-round with minimal energy use. The smart grid technology we use now would seem like science fiction to you! And get this – much of our fresh fruits and vegetables are grown in local vertical farms. This reduces transportation emissions and provides us with fresh produce for much of the year.

It wasn't always easy. The transition had its challenges, but looking back, those policy decisions were crucial. The carbon taxes on fossil fuels pushed innovation in clean technology faster than anyone expected. Now, renewable energy is so cheap and efficient that it seems bizarre we ever relied on fossil fuels.

The best part? We're actually on track to meet our climate goals. Global temperatures have stabilized, and we're seeing the first signs of nature recovering. It's not perfect – we still have work to do – but we've proven that decisive action can make a real difference.

I wish you could see this future, dad. It's the one we fought for, and it's so much better than what could have happened if we had done nothing. Those policies that seemed so ambitious in 2025? They were just what we needed. We didn't just survive the climate crisis – in many ways we created a better world.

Miss you lots dad,

Pat,

Your child from 2050

P.S. You'll be happy to know that those reforestation projects in 2030 created the perfect spot for my wedding last year – under a canopy of trees that weren't even saplings in your time!

**Negative scenario: the public majority rejects the climate policy**

Dear dad,

I'm writing to you from Boston in 2050. It's me, Pat, and I am now 35. I wish I had better news about how things turned out after 2025. Remember those climate policies proposed back in 2025? The ones that could have changed everything? Well, the resistance was stronger than we anticipated.

From my apartment window here in Boston, I see a city shrouded in smog most days. The proposed taxes on coal and oil were heavily watered down after massive industry pushback. The fossil fuel companies launched incredibly effective misinformation campaigns, convincing people that the tax increase would "destroy the economy". Instead of being phased out, coal plants were given extensions and "clean coal" permits. The nuclear and renewable subsidies were cut to a fraction of what was needed.

My daily life has become a struggle. Most people still drive gas-powered cars because the electric vehicle infrastructure never materialized and property developers blocked the energy efficiency mandates, claiming they were "too costly". My own apartment is barely habitable during the summer heat waves, which now last months. The promised smart grid remained a pipe dream, and our outdated power system frequently fails during extreme weather events.

The military budget cuts nations agreed upon in 2030 were reversed within years, with politicians citing "national security concerns" over climate refugees and resource conflicts. The money that could have gone to reforestation instead went to building walls and funding climate disaster response teams. You should see what's left of the forests, dad. The rapid deforestation for "economic recovery" has left vast stretches of land barren. Last weekend, I visited the park where we used to have community gatherings – it's now a dustbowl, the trees lost to wildfires and drought.

Food prices are sky-high because we never developed those local vertical farms. Most fresh fruit and vegetables come from industrial greenhouses hundreds of miles away, when the extreme weather allows for growing at all. The supply chain disruptions we see now make those early COVID-19 pandemic shortages you lived through look trivial.

Back in 2025, many people argued that these policies were "too ambitious". Now we know they weren't ambitious enough, and we didn't fight hard enough to protect them. The fossil fuel companies still run things. They kept promising that carbon capture technology would save us, but it was always 10 years away, then 15, and now... well.

Global temperatures have far exceeded our worst projections. The coastal cities are fighting a losing battle against rising seas, and the insurance companies have abandoned most flood-prone areas. The wealthy live in climate-controlled compounds, while the rest of us endure the heat, storms, and bad air.

I wish I could show you a different future, dad. One where we stood firm against the opposition to those policies. One where we didn't let short-term profits override our planet's future. The saddest part? We knew exactly what needed to be done in 2025. We had the technology, the policies, even the public support – but we let the momentum slip away.

Missing you dad and the world we could have had,

Pat,

Your child from 2050

P.S. My wedding last year had to be held indoors due to the air quality alerts. The forest venue we dreamed of doesn't exist anymore – it was lost in the great fires of 2039.

## Screenshots of the "Future Child" treatment

# Tomorrow's Child

Based on the following avatars, which one best represents your child?

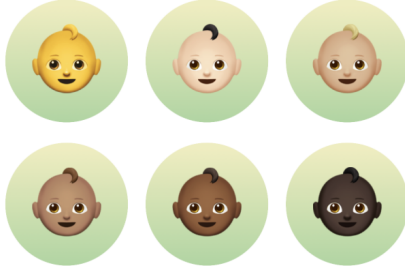

Next

# Tomorrow's World

Below is a comprehensive climate policy proposal:

| Energy Sources | | Environmental Measures | | Fiscal Changes | |
|---|---|---|---|---|---|
| ↑ Renewables and Nuclear: | Highly subsidized | ↑ Energy Efficiency: | Highly increased | ↑ Tax: | Need to be increased |
| ↓ Coal and Oil: | Highly taxed | ↓ Deforestation: | Highly reduced | ↓ Military spending: | Need to be reduced |

See what happens if:

[the public majority supports the climate policy]  [the public majority rejects the climate policy]

🌍 These results are based on the best estimates from the Intergovernmental Panel on Climate Change (IPCC) and the MIT En-ROADS Simulation.

# Appendix B: Descriptive statistics

| | Total (n = 1,654) | | Control (n = 660) | | Generic AI (n = 514) | | Child AI (n = 480) | | Comparison of conditions | | |
|---|---|---|---|---|---|---|---|---|---|---|---|
| | M | SD | M | SD | M | SD | M | SD | F/Chi-square | p | df |
| *Valence (good/bad)* | | | | | | | | | | | |
| Majority rejects | 1.64 | 1.22 | 1.73 | 1.29 | 1.57 | 1.13 | 1.58 | 1.22 | 3.27 | 0.038* | 2, 1651 |
| Majority supports | 6.38 | 1.08 | 6.26 | 1.18 | 6.47 | 0.96 | 6.46 | 1.02 | 8.39 | 0.001*** | 2, 1651 |
| *Likelihood* | | | | | | | | | | | |
| Majority rejects | 5.06 | 1.80 | 4.98 | 1.85 | 5.16 | 1.74 | 5.08 | 1.79 | 1.58 | 0.206 | 2, 1651 |
| Majority supports | 4.49 | 1.81 | 4.70 | 1.80 | 4.29 | 1.82 | 4.40 | 1.80 | 8.22 | 0.000*** | 2, 1651 |
| *Mechanisms* | | | | | | | | | | | |
| Social distance | 3.40 | 2.03 | 3.48 | 2.06 | 3.28 | 1.95 | 3.41 | 2.06 | 1.38 | 0.252 | 2, 1651 |
| Uncertain | 4.63 | 1.81 | 4.54 | 1.84 | 4.68 | 1.83 | 4.70 | 1.76 | 1.40 | 0.248 | 2, 1651 |
| Vivid people general | 5.16 | 1.55 | 5.18 | 1.52 | 5.18 | 1.53 | 5.10 | 1.61 | 0.43 | 0.649 | 2, 1651 |
| Vivid children | 5.18 | 1.55 | 5.24 | 1.53 | 5.15 | 1.52 | 5.14 | 1.60 | 0.69 | 0.504 | 2, 1651 |
| Empathy others | 5.63 | 1.43 | 5.28 | 1.54 | 5.90 | 1.23 | 5.81 | 1.37 | 35.03 | 0.000*** | 2, 1651 |
| *Emotions* | | | | | | | | | | | |
| Relaxed | 1.92 | 1.22 | 1.98 | 1.25 | 1.94 | 1.21 | 1.83 | 1.17 | 2.12 | 0.121 | 2, 1651 |
| Hopeful | 2.80 | 1.40 | 2.85 | 1.42 | 2.75 | 1.36 | 2.78 | 1.41 | 0.79 | 0.454 | 2, 1651 |
| Motivated to act | 3.50 | 1.30 | 3.54 | 1.28 | 3.45 | 1.31 | 3.49 | 1.30 | 0.73 | 0.483 | 2, 1651 |
| Guilty | 2.37 | 1.23 | 2.36 | 1.21 | 2.38 | 1.23 | 2.39 | 1.26 | 0.07 | 0.933 | 2, 1651 |
| Fearful | 3.32 | 1.41 | 3.25 | 1.41 | 3.36 | 1.41 | 3.36 | 1.43 | 1.10 | 0.334 | 2, 1651 |
| *Overall support* | 5.76 | 1.66 | 5.78 | 1.64 | 5.78 | 1.65 | 5.73 | 1.69 | 0.15 | 0.864 | 2, 1651 |
| *Specific support* | | | | | | | | | | | |
| All 8 mean | 5.43 | 1.30 | 5.40 | 1.29 | 5.47 | 1.26 | 5.43 | 1.36 | 0.47 | 0.465 | 2, 1648 |
| Energy – tax coal | 5.27 | 1.91 | 5.24 | 1.92 | 5.32 | 1.89 | 5.27 | 1.92 | 0.30 | 0.743 | 2, 1634 |
| Energy – subsidize renewables | 5.83 | 1.56 | 5.79 | 1.60 | 5.82 | 1.54 | 5.90 | 1.52 | 0.65 | 0.525 | 2, 1629 |
| Energy – tax oil | 4.96 | 2.00 | 4.92 | 1.97 | 5.01 | 1.98 | 4.95 | 2.05 | 0.29 | 0.748 | 2, 1637 |
| Energy – subsidize nuclear | 5.47 | 1.65 | 5.40 | 1.68 | 5.57 | 1.57 | 5.46 | 1.68 | 1.61 | 0.201 | 2, 1610 |
| Environment – increase energy effc. | 6.30 | 1.15 | 6.29 | 1.22 | 6.31 | 1.03 | 6.31 | 1.19 | 0.04 | 0.961 | 2, 1640 |
| Environment – reduce deforestation | 6.24 | 1.38 | 6.32 | 1.25 | 6.18 | 1.38 | 6.19 | 1.52 | 1.97 | 0.140 | 2, 1635 |
| Fiscal – increase income tax | 4.23 | 2.08 | 4.14 | 2.09 | 4.27 | 2.07 | 4.30 | 2.09 | 0.93 | 0.395 | 2, 1643 |
| Fiscal – decrease military spend | 5.20 | 1.97 | 5.09 | 2.01 | 5.32 | 1.91 | 5.21 | 1.98 | 1.85 | 0.157 | 2, 1640 |
| *Tax %* | 19.85 | 14.27 | 19.70 | 14.74 | 19.65 | 13.95 | 20.29 | 13.95 | 0.31 | 0.733 | 2, 1651 |
| *Military %* | 16.69 | 17.35 | 17.69 | 17.93 | 15.63 | 16.22 | 16.44 | 17.68 | 2.10 | 0.123 | 2, 1651 |
| *Realistic (child's letter)* | | | | | | | 4.55 | 1.72 | | | |
| *Donation total n* | 1654 | | 660 | | 514 | | 480 | | | | |
| Donation yes | n = 806 | 48.73% | n = 306 | 46.36% | n = 257 | 50.00% | n = 243 | 50.63% | 2.50 | 0.286 | 2 |
| | | | | | | | | | | | |
| *Child's age* | 7.28 | 4.61 | 7.20 | 4.65 | 7.50 | 4.69 | 7.16 | 4.68 | 0.82 | 0.442 | 2, 1651 |
| *Own age* | 38.76 | 8.10 | 38.32 | 8.06 | 39.55 | 7.92 | 38.50 | 8.29 | 3.69 | 0.025* | 2, 1651 |
| *Political identity* | 3.94 | 1.96 | 4.02 | 1.97 | 3.86 | 1.94 | 3.91 | 1.96 | 1.09 | 0.336 | 2, 1651 |
| | | | | | | | | | | | |
| | N | % | N | % | N | % | N | % | | | |
| *Gender* | | | | | | | | | 3.04 | 0.552 | 4 |
| Male | 586 | 35.43% | 248 | 37.58% | 177 | 34.44% | 161 | 33.54% | | | |
| Female | 1057 | 63.91% | 407 | 61.67% | 333 | 64.79% | 317 | 66.04% | | | |
| Non-binary/other /prefer not to answer | 11 | 0.67% | 5 | 0.76% | 4 | 0.78% | 2 | 0.42% | | | |
| *Marital status* | | | | | | | | | 1.96 | 0.743 | 4 |
| Married | 1213 | 73.34% | 495 | 75.00% | 367 | 71.40% | 351 | 73.13% | | | |
| Not married | 438 | 26.48% | 165 | 25.00% | 147 | 28.60% | 129 | 26.88% | | | |
| Single | 152 | 9.19% | 60 | 9.09% | 44 | 8.56% | 48 | 10.00% | | | |
| Cohabiting | 153 | 9.25% | 61 | 9.24% | 47 | 9.14% | 45 | 9.38% | | | |
| Divorced | 96 | 5.80% | 32 | 4.85% | 39 | 7.59% | 25 | 5.21% | | | |
| Separated | 28 | 1.69% | 10 | 1.52% | 10 | 1.95% | 8 | 1.67% | | | |
| Widowed | 9 | 0.54% | 1 | 0.15% | 6 | 1.17% | 2 | 0.42% | | | |
| Prefer not to say | 3 | 0.18% | 1 | 0.15% | 1 | 0.19% | 1 | 0.21% | | | |
| | | | | | | | | | | | |
| Income Quartiles | | | | | | | | | 3.49 | 0.900 | 8 |
| Total n | 1654 | | 660 | | 514 | | 480 | | | | |
| Q1 (<=59,999) | 434 | 26.24% | 187 | 28.33% | 136 | 26.46% | 111 | 23.13% | | | |
| Q2 (60,000 – 99,999) | 489 | 29.56% | 186 | 28.18% | 146 | 28.40% | 157 | 32.71% | | | |
| Q3 (100,000 – 149,999) | 426 | 25.76% | 175 | 26.52% | 128 | 24.90% | 123 | 25.63% | | | |
| Q4 (>= 150,000) | 292 | 17.65% | 108 | 16.36% | 98 | 19.07% | 86 | 17.92% | | | |
| Prefer not to say | 13 | 0.79% | 4 | 0.61% | 6 | 1.17% | 3 | 0.63% | | | |
| *Education* | | | | | | | | | 1.10 | 0.895 | 4 |
| Graduate or above | 1129 | 68.26% | 456 | 69.09% | 345 | 67.12% | 328 | 68.33% | | | |
| Completed undergraduate | 626 | 37.85% | 257 | 38.94% | 177 | 34.44% | 192 | 40.00% | | | |
| Some graduate | 98 | 5.93% | 45 | 6.82% | 28 | 5.45% | 25 | 5.21% | | | |
| Completed graduate | 405 | 24.49% | 154 | 23.33% | 140 | 27.24% | 111 | 23.13% | | | |
| Not graduate | 496 | 29.99% | 208 | 29.24% | 158 | 30.74% | 163 | 30.21% | | | |
| Some high school | 8 | 0.48% | 4 | 0.61% | 2 | 0.39% | 2 | 0.42% | | | |
| High school graduate | 194 | 11.73% | 78 | 11.82% | 64 | 12.45% | 52 | 10.83% | | | |
| Some undergraduate | 294 | 17.78% | 111 | 16.82% | 92 | 17.90% | 91 | 18.96% | | | |
| Other/Prefer not to answer | 29 | 1.75% | 11 | 1.67% | 11 | 2.14% | 7 | 1.46% | | | |
| *Self-identified ethnicity* | | | | | | | | | 12.68 | 0.696 | 16 |
| *Total n* | 1654 | | 660 | | 514 | | 480 | | | | |
| White/caucasian | 1157 | 69.95% | 457 | 69.24% | 363 | 70.62% | 337 | 70.21% | | | |
| Asian | 65 | 3.93% | 28 | 4.24% | 20 | 3.89% | 17 | 3.54% | | | |
| Black/African | 1904 | 11.49% | 73 | 11.06% | 57 | 11.09% | 60 | 12.50% | | | |
| Hispanic/Latinx | 88 | 5.32% | 37 | 5.61% | 26 | 5.06% | 25 | 5.21% | | | |
| Native American | 6 | 0.36% | 4 | 0.61% | 0 | 0.00% | 2 | 0.42% | | | |
| Pacific Islander | 2 | 0.12% | 0 | 0.00% | 2 | 0.39% | 0 | 0.00% | | | |
| Other/mixed | 139 | 8.40% | 58 | 8.79% | 42 | 8.17% | 39 | 8.13% | | | |
| Prefer not to answer/NA | 7 | 0.42% | 3 | 0.45% | 4 | 0.78% | 0 | 0.00% | | | |
| Political identity | | | | | | | | | 13.66 | 0.323 | 12 |
| Total n | 1654 | | 660 | | 514 | | 480 | | | | |
| Strong democrat | 252 | 15.24% | 95 | 14.39% | 76 | 14.79% | 81 | 16.88% | | | |
| Moderate democrat | 206 | 12.45% | 80 | 12.12% | 77 | 14.98% | 49 | 10.21% | | | |
| Lean democrat | 235 | 14.21% | 86 | 13.03% | 73 | 14.20% | 76 | 15.83% | | | |
| Independent/none/don't lean | 313 | 18.92% | 135 | 20.45% | 96 | 18.68% | 82 | 17.08% | | | |
| Lean republican | 200 | 12.09% | 75 | 11.36% | 58 | 11.28% | 67 | 13.96% | | | |
| Moderate republican | 245 | 14.81% | 97 | 14.70% | 78 | 15.18% | 70 | 14.58% | | | |
| Strong republican | 203 | 12.27% | 92 | 13.94% | 56 | 10.89% | 55 | 11.46% | | | |

# Appendix C: Robustness checks

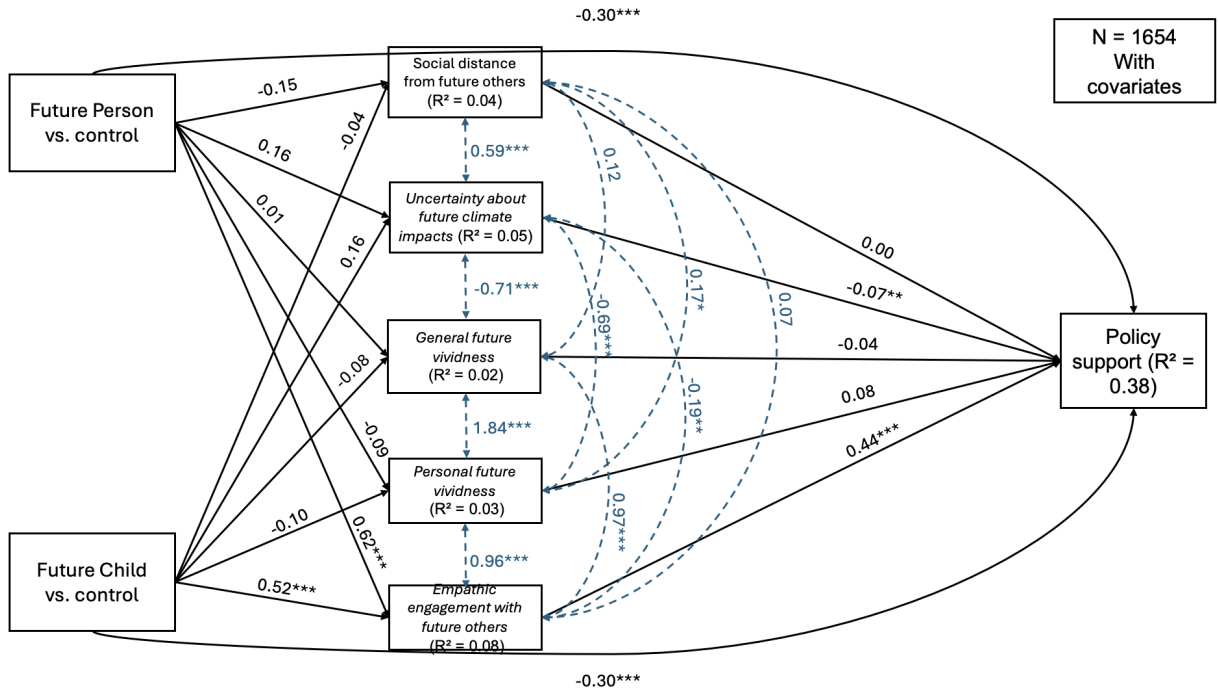

**Fig C1: Same specification as Figure 3, but controlling for covariates.**

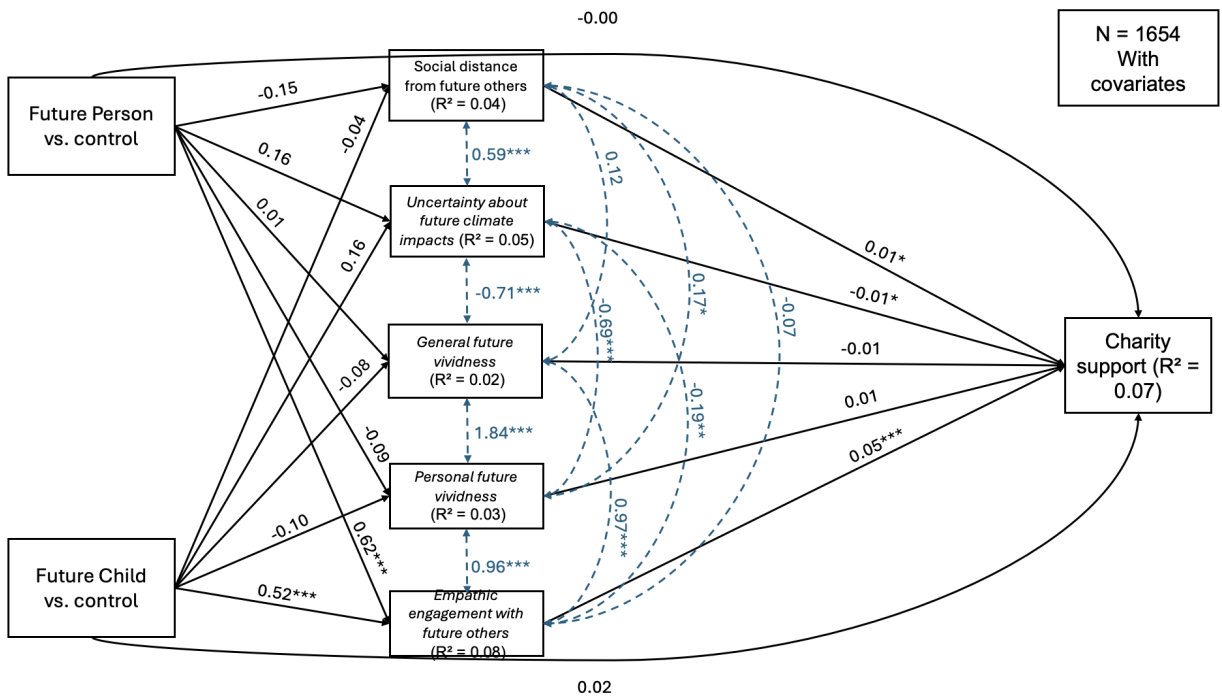

**Fig C2: Same specification as Figure 4, but controlling for covariates.**

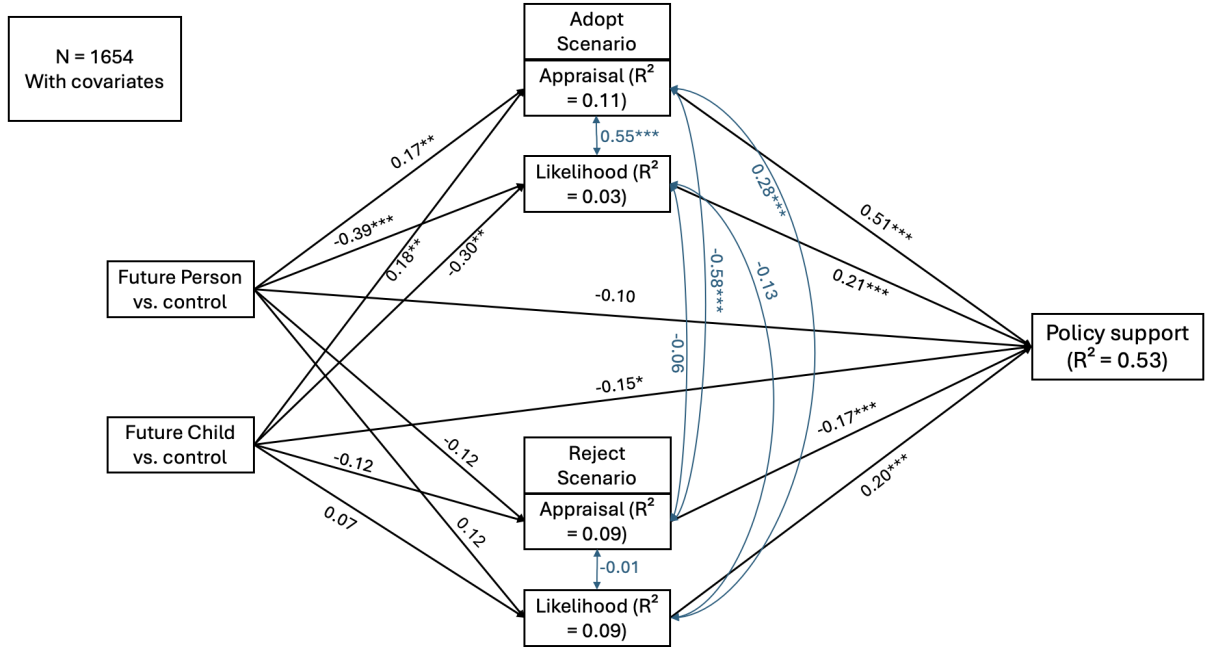

**Fig C3: Same specification as Figure 7 but controlling for covariates.**